\documentclass[sigconf,screen]{acmart}

\usepackage{booktabs}
\usepackage{amsmath}
\usepackage{graphicx}
\usepackage{subcaption}
\usepackage{algorithm}
\usepackage{float}                 
\usepackage{enumitem}
\usepackage{balance}
\usepackage{xspace}
\usepackage{multirow}
\usepackage{microtype}
\usepackage{mathtools}
\usepackage{soul}
\sethlcolor{red!20}
\usepackage{hyperref}

\newcommand{\rqbox}[1]{%
  \par\noindent
  \begingroup
    \setlength{\fboxrule}{0pt}
    \setlength{\fboxsep}{2pt}
    \colorbox{gray!10}{%
      \parbox{\dimexpr\linewidth-2\fboxsep\relax}{%
        \vspace{0.5pt}
        #1%
      }%
    }%
  \endgroup
  \par
}

\newboolean{showcomments}
\setboolean{showcomments}{true}
\ifthenelse{\boolean{showcomments}}
{
	\definecolor{myyellow}{RGB}{255, 228, 26}
	\definecolor{myblue}{RGB}{50, 50, 220}
	\newcommand{\nb}[2]{
		{\sf
			\fcolorbox{myyellow}{yellow}{\scriptsize\textbf{#1}}%
			$\blacktriangleright$%
			{\color{myblue}\fontsize{7pt}{8pt}\selectfont\textbf{#2}}%
		}%
	}
}
{
	\newcommand{\nb}[2]{}
}

\renewcommand*{\equationautorefname}{Equation}
\def\equationautorefname~#1\null{(#1)\null}

\newcommand{\lpips}{\textsc{LPIPS}\xspace}
\newcommand{\psiT}{\ensuremath{\psi}\xspace}
\newcommand{\psistar}{\ensuremath{\psi^\star}\xspace} 
\newcommand{\datasetA}{MNIST\xspace}
\newcommand{\datasetB}{Fashion MNIST\xspace}
\newcommand{\datasetC}{CIFAR-10\xspace}

\newcommand{\Wplus}{\mathcal{W}^+}


\title{Latent Regularization in Generative Test Input Generation}

\author{Giorgi Merabishvili}
\orcid{0009-0000-0314-7487}
\email{gmerabi@ncsu.edu}
\affiliation{%
  \institution{North Carolina State University}
  \country{USA}
}

\author{Oliver Wei{\ss}l}
\orcid{0009-0008-7575-0187}
\email{weissl@fortiss.org}
\affiliation{%
  \institution{Technical University of Munich \& fortiss GmbH}
  \country{Germany}
}

\author{Andrea Stocco}
\orcid{0000-0001-8956-3894}
\email{andrea.stocco@tum.de}
\affiliation{%
  \institution{Technical University of Munich \& fortiss GmbH}
  \country{Germany}
}

\begin{document}

\begin{abstract}
This study investigates the impact of regularization of latent spaces through truncation on the quality of generated test inputs for deep learning classifiers. We evaluate this effect using style-based GANs, a state-of-the-art generative approach, and assess quality along three dimensions: validity, diversity, and fault detection.
We evaluate our approach on the boundary testing of deep learning image classifiers across three datasets, \datasetA, \datasetB, and \datasetC.
We compare two truncation strategies: latent code mixing with binary search optimization and random latent truncation for generative exploration.
Our experiments show that the latent code-mixing approach yields a higher fault detection rate than random truncation, while also improving both diversity and validity.
\end{abstract}

\copyrightyear{2026}
\acmYear{2026}
\setcopyright{cc}
\setcctype{by}
\acmConference[DeepTest '26]{7th International Workshop on Deep Learning for Testing and Testing for Deep Learning (DeepTest '26)}{April 12--18, 2026}{Rio de Janeiro, Brazil}
\acmBooktitle{7th International Workshop on Deep Learning for Testing and Testing for Deep Learning (DeepTest '26) (DeepTest '26), April 12--18, 2026, Rio de Janeiro, Brazil}
\acmPrice{}
\acmDOI{10.1145/3786154.3788585}
\acmISBN{979-8-4007-2386-5/2026/04}

\maketitle

\section{Introduction}\label{sec:intro}

Testing deep learning (DL) systems requires large, diverse, and valid inputs to expose misbehaviors~\cite{pei2017deepxplore,tian2018deeptest,ma2018deepgauge,guo2019dlfuzz,xie2019deephunter,odena2019tensorfuzz,riccio2020testing,2020-Humbatova-ICSE,2022-Stocco-ASE,2022-Stocco-TSE,2020-Stocco-ICSE,2023-Stocco-EMSE,2021-Stocco-JSEP,2024-Grewal-ICST,lambertenghi2024assessing,2025-Lambertenghi-ASE,2025-Lambertenghi-ICST,2020-Stocco-GAUSS}.
As deep learning models and datasets grow in complexity, manually creating such inputs becomes infeasible due to the huge size of the input space. Test input generators can be used to sample from this space more effectively. Recent generative AI methods, such as style-based GANs, are promising as they produce high-quality synthetic data and enable structured manipulations in their latent spaces~\cite{goodfellow2014gan,karras2019stylegan,karras2021aliasfree}.
Generating test inputs requires balancing two conflicting objectives: the inputs should be valid (plausible and human-recognizable) but also diverse enough to explore the DL system decision space and reveal faults. 

Previous work has used GANs to generate test inputs with semantic variations (for example, weather changes in autonomous driving)~\cite{zhang2018deeproad}, using mechanisms such as \emph{truncation} to balance realism and diversity~\cite{brock2019large,maryam2025benchmarking}. 
Truncation works by moving latent codes toward their mean, producing more realistic but less diverse images.
Although truncation is part of most modern GANs~\cite{sauer2022styleganxl}, its effect on generating test inputs for DL systems has not been thoroughly analyzed.
Existing DL testing work usually relies on pixel or feature perturbations~\cite{pei2017deepxplore,guo2019dlfuzz}, or explores semantic changes without considering truncation as a controllable factor~\cite{tian2018deeptest,zhang2018deeproad}.
Recent search-based approaches~\cite{riccio2020deepjanus,kang2020sinvad,zohdinasab2021deephyperion,zohdinasab2023deephyperioncs, weissl2025targeted, dola2024cit4dnn} explore other forms of variability, but not this specific trade-off.

This work explores how truncation, a common latent-space regularization method, influences the quality of test inputs produced by class-conditional StyleGAN2 models. These models form the foundation of recent generative testing frameworks, such as \textsc{Mimicry}~\cite{weissl2025targeted}, which we adopt as a reference in this study.
We first perform a screening step on the generated input seeds to filter out invalid samples, using a combination of classifier confidence, output margin, SSIM, and $\ell_2$ threshold~\cite{heusel2017ttur,salimans2016improved,sajjadi2018pr,kynkaanniemi2019precisionrecall}. 
We further evaluate an adaptive truncation strategy, in which we progressively decrease the truncation rate until valid candidate seeds are produced. These ``salvaged'' seeds typically lie close to the classifier's decision boundaries and are therefore more likely to trigger misclassifications. Using these seeds, we then generate failure-inducing inputs with Mimicry through a search-based test generation procedure.

We conduct experiments on three datasets, namely \datasetA, \datasetB, and \datasetC, varying the truncation level to study its effect on multiple quality metrics:
(i)~the proportion of samples passing automatic screening,
(ii)~the fraction of human-validated valid samples,
(iii)~the diversity among validated samples, and
(iv)~the fault detection rate among validated inputs.

Our experiments demonstrate the advantage of the adaptive seed selection strategy. In the \datasetA\ benchmark targeting 25 frontier pairs, 15 samples were confirmed as valid and fault-revealing, with 12 of them originating from salvaged seeds.
Overall, our results indicate that adaptive truncation improves the acceptance of generated test inputs while maintaining high diversity and increasing fault detection effectiveness.

Our experiment revealed that mild–moderate latent regularization (\mbox{$\psiT{\approx}0.6$}) maximizes human-validated frontier yield on \datasetA. The adaptive policy minimizes human effort (seeds/valid) by salvaging hard seeds. Qualitative analysis shows that diversity remains adequate at these settings to expose faults, whereas overly strong regularization offers lower returns.
First-flip is the most efficient for fault discovery under latent regularization; \emph{style-mixing} complements it by adding semantic control over boundary direction. Using first-flip to triage seeds (possibly with adaptive $\psiT$) and \emph{style-mixing} provides strong overall coverage of boundary behaviors.

Our paper makes the following contributions:

\begin{description}[noitemsep]

\item [Systematic study of truncation for test generation.] We conduct the first structured investigation of how latent-space regularization through truncation affects the quality of generated test inputs for deep learning classifiers.

\item [Adaptive truncation strategy for seed selection.] We propose an adaptive truncation method that progressively lowers the truncation parameter $\psiT$ to salvage seeds near decision boundaries, improving fault discovery and diversity.

\item [Experimental evaluation.] We evaluate both random and search-based truncation strategies on three datasets, \datasetA, \datasetB, and \datasetC, measuring validity, diversity, and fault detection under different truncation levels. Our experiments show that adaptive truncation improves the acceptance rate and fault detection capability of generated test inputs, offering practical guidance for designing generative testing frameworks.
\end{description}

\section{Background}\label{sec:background}

\subsection{DL Testing Objectives}

Testing of DL systems generally serves two goals: 
\emph{robustness testing}, which stresses a trained model using semantically preserved perturbations, and \emph{generalization testing}, which seeks novel yet \emph{valid} inputs that traverse or expose decision boundaries~\cite{pei2017deepxplore,tian2018deeptest,ma2018deepgauge,guo2019dlfuzz,xie2019deephunter}.

Boundary testing is a form of generalization testing that focuses on regions of near-ambiguity~\cite{10.1007/s10664-023-10393-w}, where class confidences are similar. It can be performed in an \emph{untargeted} way (searching for any label flip) or \emph{targeted} (moving from a source class toward a specific rival suggested by the model's confidence distribution). Targeted exploration is usually more sample-efficient in high-dimensional spaces, as it constrains the search to a source–rival corridor on the decision surface~\cite{moosavi2016deepfool,brendel2018boundary,chen2020hopskipjump, weissl2025targeted}. 

\subsection{Test Input Generation Families}

Existing DL testing approaches can be broadly grouped into input-based, model-based, and latent-space methods. Input-based techniques perturb existing samples directly in the input space, for example, via gradient-based approaches such as DeepXplore~\cite{pei2017deepxplore} and DLFuzz~\cite{guo2019dlfuzz}, or random and search-based transformations~\cite{tian2018deeptest}. While simple and model-agnostic, these methods mainly target robustness testing and offer limited control over semantic validity. Model-based methods instead explore explicit domain abstractions to generate challenging yet valid inputs~\cite{riccio2020deepjanus,zohdinasab2021deephyperion,zohdinasab2023deephyperioncs,2025-Chen-EMSE}, but their effectiveness depends on the fidelity of the underlying model.

Latent-space methods aim to provide realistic inputs using a generative model and explore its latent space to produce diverse, semantically plausible test cases~\cite{kang2020sinvad,weissl2025targeted,2025-Guo-arxiv}. Recent GenAI-based frameworks, including \textsc{Mimicry}~\cite{weissl2025targeted}, \textsc{RBT4DNN}~\cite{mozumder2025rbt4dnn}, \textsc{CIT4DNN}~\cite{dola2024cit4dnn}, and \textsc{GIFTbench}~\cite{maryam2025benchmarking}, use different forms of latent-space exploration, including interpolation, noise injection, and guided search, to generate semantically valid and behavior-revealing test inputs.

\begin{figure}[t]
  \centering
  \setlength{\tabcolsep}{1pt}
  \renewcommand{\arraystretch}{0}
  \scriptsize
  \resizebox{0.5\textwidth}{!}{%
    \begin{tabular}{c c c c c c}
    $\psiT{=}1.0$ & $0.8$ & $0.6$ & $0.4$ & $0.2$ \\[2pt]
    \includegraphics[width=0.04\textwidth]{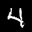} &
    \includegraphics[width=0.04\textwidth]{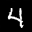} &
    \includegraphics[width=0.04\textwidth]{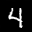} &
    \includegraphics[width=0.04\textwidth]{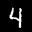} &
    \includegraphics[width=0.04\textwidth]{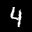} &
    \\[4pt]

    \includegraphics[width=0.04\textwidth]{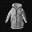} &
    \includegraphics[width=0.04\textwidth]{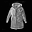} &
    \includegraphics[width=0.04\textwidth]{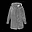} &
    \includegraphics[width=0.04\textwidth]{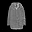} &
    \includegraphics[width=0.04\textwidth]{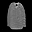} &
    \\[4pt]

    \includegraphics[width=0.04\textwidth]{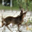} &
    \includegraphics[width=0.04\textwidth]{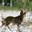} &
    \includegraphics[width=0.04\textwidth]{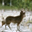} &
    \includegraphics[width=0.04\textwidth]{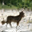} &
    \includegraphics[width=0.04\textwidth]{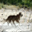} &
    \\
  \end{tabular}
  }
  \caption{\textbf{Truncation on coarse layers. A fixed latent seed at $\psi_T=1.0$ and progressively lower $\psi_T$ values increase fidelity and reduce diversity.}}
  \label{fig:trunc_sweep}
\end{figure}


In this work, we focus on latent-space methods and isolate the role of a single, widely used control parameter—\emph{truncation}—as a factor influencing the validity, diversity, and fault-revealing power of generated test inputs. Our study builds on \textsc{Mimicry}, which serves as the reference generator for all experiments.

\textsc{Mimicry} uses a class-conditional StyleGAN to structure the latent space into an intermediate representation $\mathcal{W}$, obtained by mapping an initial latent vector $z \in \mathcal{Z}$. This representation supports layer-wise control and semantic manipulation via style or feature mixing~\cite{karras2019stylegan,harkonen2020ganspace,shen2021sefa,bau2019gandissect,weissl2025targeted}. Test generation is formulated as a search for \emph{boundary inputs}, i.e., samples located near the decision boundary of the system under test, where class confidences are balanced. We leverage this framework to study how truncation reshapes the latent space during boundary exploration.

\section{Methodology}\label{sec:method}

In generative test input generators such as \textsc{Mimicry}, the truncation parameter $\psi$ can be used as a regularization control that moves each latent code toward the average latent $\bar{w}$, trading diversity for realism. 
The so-called \emph{truncation trick} contracts latent codes $w$ toward the running average $\bar{w}$ according to:
\[
w' = \bar{w} + \psiT (w - \bar{w}), \qquad \psiT \in (0,1].
\]
Lower values of $\psiT$ typically increase visual realism and classifier confidence but reduce the diversity of generated samples~\cite{brock2019large,sauer2022styleganxl}. This makes truncation a natural variable for studying how fidelity controls influence the generation of valid and fault-revealing test inputs. However, test input generation must also balance other competing goals: \emph{validity}, i.e., producing class-plausible images, and \emph{diversity}, i.e., covering enough of the input space to traverse decision boundaries and expose faults. 

Truncation in StyleGAN can be applied uniformly across all layers in the latent code $w$ or on a per-layer basis in $\Wplus$~\cite{}. Uniform schedules simplify comparisons, while layer-wise schedules allow different truncation values per layer to preserve coarse semantic structure~\cite{karras2019stylegan,bau2019gandissect,shen2021sefa}. Lowering the truncation parameter $\psiT$ concentrates samples in high-density regions, typically increasing fidelity and classifier confidence while reducing diversity.
An example is given in \autoref{fig:trunc_sweep}.

\begin{figure}[t]
  \centering
  \includegraphics[width=0.9\columnwidth]{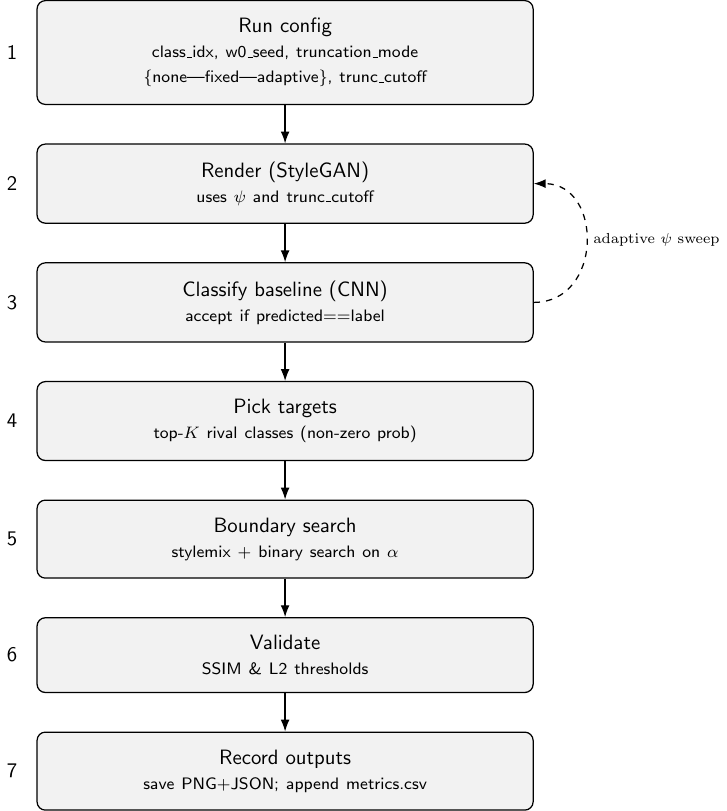}
  \caption{Test input generation with truncation workflow.}
  \label{fig:workflow}
\end{figure}

\subsection{Approach}\label{sec:approach}

We use two key notions throughout our approach. The \emph{minimal acceptable truncation} $\psistar$ is the smallest $\psiT$ at which a seed remains behavior-preserving for the system under test (SUT); this forms the basis of \emph{seed salvage}. The \emph{first-flip truncation}, instead, is the first $\psiT$ (descending from $1.0$) at which the SUT prediction changes.

Our test input generation workflow is illustrated in \autoref{fig:workflow}. Each run is configured with a class index, latent seed, truncation mode (\emph{none, fixed, adaptive}), and truncation cutoff, specifying the feature layer up until truncation will be applied, starting from the bottom, giving an option to truncate only lower layers and leave fine details unchanged.
Images are rendered at the chosen $\psiT$ with the initial latent $z$ frozen to isolate truncation effects~\cite{}. The baseline check at $\psiT=1.0$ ensures that only seeds passing the fixed confidence and margin requirements are retained. For boundary testing, we select the top-$K$ rival classes according to the DL classifier probabilities whenever we have non-zero probabilities for classes other than the source~\cite{riccio2020deepjanus}. Each candidate then undergoes boundary exploration using one of two techniques: truncation-assisted style mixing or truncation-only first-flip truncation. After the search, candidates are evaluated based on SSIM and $\ell_2$ thresholds to filter out invalid frontier pairs early on.

In \emph{truncation-assisted style mixing} (Algorithm~\ref{alg:stylemix}), for each seed $z$, we first render the untruncated baseline $x_{1.0}=G(z,c;1.0)$ and select the top-$K$ rival classes $R$ by classifier confidence. For each truncation budget $b \in B$, we then render a truncated source image $x_b = G(z,c;b)$ and, for each rival $r \in R$, style-mix the source-class latent $w_c^{(b)}$ with the rival latent $w_r^{(b)}$ over a specified set of layers $L$~\cite{karras2019stylegan}. We run a binary search over the mixing weight $\lambda \in [0,1]$ to find the smallest $\lambda$ that flips the SUT prediction from the source class while keeping the similarity constraints (SSIM and $\ell_2$). 

\begin{algorithm}[t]
\caption{Testing via Truncation-Assisted Style Mixing}
\label{alg:stylemix}
\small
\noindent\textbf{Input.}
Generator $G$, classifier $h$; truncation budgets $B$ ($\psi_T$);
seeds $\mathcal{Z}$; top-$K$ rivals; layers $L$; binary-search steps $T$;
SSIM and $\ell_2$ threshold $\tau$.

\noindent\textbf{Output.}
A set of recorded frontiers with metadata $(x_b, x_\lambda, b, \lambda)$.

\begin{enumerate}
\item For each seed $z \in \mathcal{Z}$:
  \begin{enumerate}
  \item Render baseline: $x_{1.0} \leftarrow G(z, c; 1.0)$.
  \item Let $R \leftarrow$ the top-$K$ rival classes by confidence on $x_{1.0}$.
  \item For each truncation budget $b \in B$:
    \begin{enumerate}
    \item Render truncated source: $x_b \leftarrow G(z, c; b)$.
    \item For each rival $r \in R$:
      \begin{enumerate}
      \item Run a binary search over $\lambda \in [0,1]$ (for $T$ steps). At each step,
            generate $x_\lambda \leftarrow \text{StyleMix}(w_c^{(b)}, w_r^{(b)}, L, \lambda)$.
      \item If $x_\lambda$ satisfies the similarity constraints (SSIM and $\ell_2$) w.r.t.\ $x_b$
            and $\arg\max h(x_\lambda) \ne c$, then record the frontier
            $(x_b, x_\lambda, b, \lambda)$ and stop iterating rivals for this $(z,b)$.
      \end{enumerate}
    \end{enumerate}
  \end{enumerate}
\item Aggregate all recorded frontiers and store the corresponding JSON files.
\end{enumerate}
\end{algorithm}

\begin{figure}[t]
  \centering
  \begin{subfigure}{.38\columnwidth}
    \includegraphics[width=\linewidth]{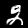}
    \caption{\scriptsize Baseline ($\psiT{=}1.00$): human-invalid, SUT passing.}
  \end{subfigure}
  \hspace{4mm}
  \begin{subfigure}{.38\columnwidth}
    \includegraphics[width=\linewidth]{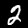}
    \caption{\scriptsize Truncated ($\psiT{=}0.55$): human-valid,  SUT failing.}
  \end{subfigure}
  \caption{Truncation-only search. Truncation refines a human-invalid baseline into a human-valid image and flips the SUT prediction.}
  \label{fig:interesting_case}
\end{figure}

In \emph{truncation-only first-flip} (Algorithm~\ref{alg:firstflip}), for each seed $z$ we first generate a baseline image $x_{1.0}=G(z,c;1.0)$. We then decrease the truncation parameter $\psiT$ along a predefined descending schedule and re-render the image at each step. The search stops at the first $\psi^\star$ for which the classifier prediction flips from the source class, and the corresponding truncated image is recorded as a frontier case. \autoref{fig:interesting_case} illustrates a truncation-only refinement: a baseline sample that is human-invalid at $\psi{=}1.00$ becomes human-valid at $\psi{=}0.55$ and simultaneously flips to a rival class. This example shows how moderate truncation can raise validity while moving the sample closer to a decision boundary, even without style mixing.

\begin{algorithm}[t]
\caption{Truncation-Only First-Flip Search}
\label{alg:firstflip}
\small

\noindent\textbf{Input.}
Generator $G$; classifier $h$; class $c$;
schedule $\Psi = \{1.0, \psi_2, \ldots, \psi_M\}$ (descending);
seeds $\mathcal{Z}$.

\noindent\textbf{Output.}
Recorded frontiers $(z, c, x_{\psi^\star})$ with metadata.

\begin{enumerate}
\item For each seed $z \in \mathcal{Z}$:
  \begin{enumerate}
  \item Generate baseline input:
        $x_{1.0} \leftarrow G(z, c; 1.0)$.
  \item If $\arg\max h(x_{1.0}) \ne c$, skip this seed.
  \item For each truncation level $\psi_T \in \Psi \setminus \{1.0\}$ (in descending order):
    \begin{enumerate}
    \item Generate truncated image:
          $x_{\psi_T} \leftarrow G(z, c; \psi_T)$.
    \item If $\arg\max h(x_{\psi_T}) \ne c$:
      \begin{enumerate}
      \item Set $\psi^\star \leftarrow \psi_T$.
      \item Record frontier $(z, c, x_{\psi_T})$ and $\psi^\star$.
      \item Stop iterating truncation levels for this seed.
      \end{enumerate}
    \end{enumerate}
  \end{enumerate}
\item Aggregate all recorded frontiers and export the corresponding JSON files.
\end{enumerate}
\end{algorithm}


\begin{figure}[t]
  \centering
  \includegraphics[width=1.00\linewidth]{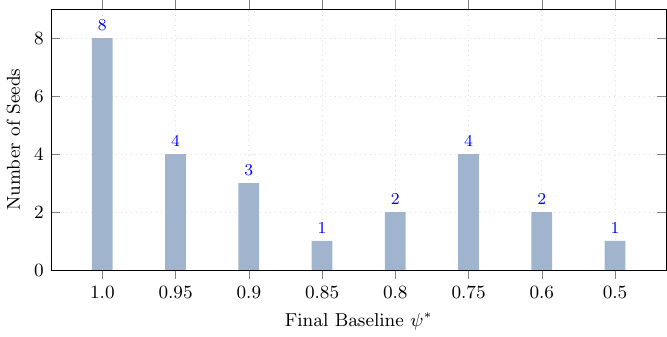}
  \caption{Minimal truncation $\psistar$ in MNIST ($n=25$).}
  \label{fig:adaptive-psi-dist}
\end{figure}

\begin{figure}[t]
  \centering
  \begin{subfigure}{.30\columnwidth}
    \includegraphics[width=\linewidth]{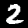}
    \caption{\scriptsize Baseline ($\psiT{=}1.00$): human-valid, SUT passing.}
  \end{subfigure}
  \hspace{4mm}
  \begin{subfigure}{.30\columnwidth}
    \includegraphics[width=\linewidth]{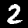}
    \caption{\scriptsize Truncated ($\psiT{=}0.75$): SUT classifies \emph{9}.}
  \end{subfigure}
  \caption{Target revelation via truncation. Lowering $\psiT$ causes human-valid seed to flip under truncation alone.}
  \label{fig:target_revelation_case}
\end{figure}

Lowering $\psiT$ can also \emph{salvage seeds} that fail the baseline at $\psiT=1.0$, moving them toward regions where they change the SUTs behavior and remain within the perceptual threshold. These salvaged seeds tend to lie closer to decision boundaries, increasing the likelihood of flips. Additionally, truncation reshapes the classifier's confidence landscape, sometimes surfacing rival classes that were suppressed at $\psiT=1.0$, a phenomenon we term \emph{target revelation}. For example, \autoref{fig:adaptive-psi-dist} shows the distribution of $\psistar$ across human-validated, fault-revealing seeds in MNIST, one of the datasets of our study. \autoref{fig:target_revelation_case} illustrates this effect, where decreasing $\psi$ affects the rival confidence and enables a human-valid seed to flip under truncation alone.

\section{Empirical Study}
\label{sec:results}

\subsection{Research Questions}
\label{sec:rqs}


\textbf{RQ\textsubscript{1} (effectiveness).} \textit{How does truncation affect the validity, diversity, and fault detection of generated test inputs?}  

    This question evaluates the core trade-off introduced by truncation. Lower truncation levels increase image realism and classifier confidence but may reduce diversity, potentially limiting fault discovery. Understanding this balance is important to selecting truncation settings that maximize both test validity and effectiveness.

\textbf{RQ\textsubscript{2} (comparison).} \textit{How do the style-mixing and first-flip methods compare in terms of efficiency and fault detection?}  

    This question contrasts two strategies for exploring the latent space: (i) \emph{style-mixing}, which searches along semantic directions via layer-wise interpolation, and (ii) \emph{first-flip}, which relies on truncation descent. Comparing their performance clarifies whether structured latent manipulations are necessary or if truncation alone suffices for generating meaningful, fault-revealing inputs.

\subsection{Experimental Setup, Metrics, and Procedure}
\label{sec:study}
\subsubsection{Objects of Study.}
We evaluate the proposed approach on three benchmark datasets: \datasetA\ (28$\times$28 grayscale)~\cite{lecun1998gradient}, \datasetB\ (28$\times$28 grayscale)~\cite{xiao2017fashionmnist}, and \datasetC\ (32$\times$32 color)~\cite{krizhevsky2009learning}.  
For each dataset, we employ pretrained class-conditional StyleGAN2(-ADA) generators as test input producers from existing work~\cite{weissl2025targeted}. The corresponding SUTs used are Torchvision baselines: a small CNN for \datasetA\ and \datasetB, and a ResNet-18 for \datasetC~\cite{he2016resnet}, also from existing work~\cite{weissl2025targeted}.

Truncation is explored over fixed budgets \autoref{eq:fb}, and an adaptive schedule \autoref{eq:ab} is used to automatically locate the minimal truncation value $\psistar$ that produces valid candidates. 

In \autoref{tab:full-results}, experiments on \datasetA, \datasetB, and \datasetC show that diversity (mean pairwise \lpips) dropped sharply at $\psiT \in \{0.6, 0.5\}$, and lower values would further reduce diversity without supporting our goal of balancing validity and diversity.
For this reason, we set a minimal truncation value at $\psiT=0.5$ in the main experiments.
\begin{align}
\psi_{\text{fixed}}  &= \{1.0,\, 0.9,\, 0.8,\, 0.7,\, 0.6,\, 0.5\} \label{eq:fb}\\
\psi_{\text{adapt}} &= \{1.0,\, 0.95,\, 0.90,\, 0.85,\, 0.80,\, 0.75,\, 0.70,\, 0.60,\, 0.50\} \label{eq:ab}
\end{align}

Each generator is evaluated with 20 latent seeds per class for \datasetA\ and \datasetB, and 30 seeds per class for \datasetC. Identical seed lists are reused across budgets for comparability. 


\begin{table*}[t]
  \centering
  \normalsize
  \caption{Results across datasets, truncation levels, and probing techniques.
  Each entry reports the number of sampled seeds, human-validated inputs, 
  validation rate, mean \lpips, fault detection rate, and 
  efficiency (seeds per validated input).
  }
  \label{tab:full-results}
  \setlength{\tabcolsep}{9.32pt}
  \renewcommand{\arraystretch}{1}
  \begin{tabular}{lllrrrrr}
    \toprule
    Dataset & Technique & Setting & Seeds & Human-val & Human-Val.\ Rate & \lpips & Fault\ Rate \\
    \midrule

    \multirow{8}{*}{\datasetA}
      & \multirow{7}{*}{Style-Mixing}
        & No trunc.     & 504 & 11 & 44\% & 0.194 & 2.18\% \\
      & & $\psiT{=}0.9$ & 518 & 12 & 48\% & 0.224 & 2.31\% \\
      & & $\psiT{=}0.8$ & 461 & 16 & 64\% & 0.215 & 3.47\% \\
      & & $\psiT{=}0.7$ & 549 & 14 & 56\% & 0.190 & 2.55\% \\
      & & $\psiT{=}0.6$ & 455 & \textbf{20} & \textbf{80\%} & 0.152 & 4.39\% \\
      & & $\psiT{=}0.5$ & 448 & 17 & 68\% & 0.143 & 3.79\% \\
      & & Adaptive      & \textbf{161} & 15 & 60\% & \textbf{0.252} & \textbf{9.31\%} \\[2pt]
      \cmidrule{2-8}
      & \multirow{1}{*}{Truncation-Only}
        & Gradual trunc.     & 499 & 19 & 76\% & 0.156 & 3.80\% \\

    \midrule
    \multirow{8}{*}{\datasetB}
      & \multirow{7}{*}{Style-Mixing}
        & No trunc.     & 431 & 23 & 92\% & \textbf{0.240} & 5.33\% \\
      & & $\psiT{=}0.9$ & 502 & 24 & 96\% & 0.236 & 4.78\% \\
      & & $\psiT{=}0.8$ & 718 & \textbf{25} & \textbf{100\%} & 0.173 & 3.48\% \\
      & & $\psiT{=}0.7$ & 706 & \textbf{25} & \textbf{100\%} & 0.203 & 3.54\% \\
      & & $\psiT{=}0.6$ & 392 & 24 & 96\% & 0.166 & 6.12\% \\
      & & $\psiT{=}0.5$ & 526 & \textbf{25} & \textbf{100\%} & 0.170 & 4.75\% \\
      & & Adaptive      & \textbf{169} & 22 & 88\% & 0.214 & \textbf{13.0\%} \\[2pt]
      \cmidrule{2-8}
      & \multirow{1}{*}{Truncation-only}
        & Gradual trunc.     & 93 & 25 & 100\% & 0.227 & 26.8\% \\

    \midrule
    \multirow{8}{*}{\datasetC}
      & \multirow{7}{*}{Style-Mixing}
        & No trunc.     & 275 & 13 & 52\% & \textbf{0.424} & 4.72\% \\
      & & $\psiT{=}0.9$ & 249 & 13 & 52\% & 0.404 & 5.22\% \\
      & & $\psiT{=}0.8$ & 252 & 16 & 64\% & 0.400 & 6.34\% \\
      & & $\psiT{=}0.7$ & 249 & 15 & 60\% & 0.406 & 6.02\% \\
      & & $\psiT{=}0.6$ & 158 & \textbf{17} & \textbf{68}\% & 0.383 & 10.7\% \\
      & & $\psiT{=}0.5$ & \textbf{118} & 16 & 64\% & 0.338 & \textbf{13.5}\% \\
      & & Adaptive      & 255 & 13 & 52\% & 0.421 & 5.09\% \\[2pt]
      \cmidrule{2-8}
      & \multirow{1}{*}{Truncation-only}
        & Gradual trunc.     & 266 & 16 & 64\% & 0.386 & 6.01\% \\
    \bottomrule
  \end{tabular}
\end{table*}

\subsubsection{Evaluation Metrics.}
We assess the impact of truncation using four operational metrics that jointly capture the realism, diversity, and fault-revealing power of generated test inputs:

\begin{itemize}[leftmargin=1.1em]
  \item \textbf{Baseline Acceptance.}  
  For a seed $z$ and class $c$, the baseline image $x_{1.0} = G(z, c; 1.0)$ is accepted if it meets three classifier-based criteria: 
  (1) $\arg\max h(x_{1.0}) = c$ (predicted as the correct class);  
  (2) top-class confidence $p_{(1)} \ge p_{\min}$;  
  (3) confidence margin $p_{(1)} - p_{(2)} \ge \delta$.  

  \item \textbf{Screening Validity.}  
  For each truncation level $b$, the generated image $x_b = G(z, c; b)$ must pass both the classifier gate above and a threshold of SSIM = 0.95 and $\ell_2$ = 0.2 that is fixed for every dataset. This step filters out visually implausible or off-mode samples before human inspection~\cite{weissl2025targeted}. 

  \item \textbf{Human-Validated Validity.}  
  Screened images are independently reviewed by human annotators to confirm whether they are class-plausible. This step captures true semantic validity that cannot be guaranteed by automated proxies.

  \item \textbf{Diversity and Fault Detection.}  
  Diversity is quantified as the mean pairwise perceptual distance (\lpips) among all human-validated samples for a given $(c, b)$, measuring coverage of the visual input space.  
      Fault detection is the proportion of validated samples that cause a misclassification, i.e., $\arg\max h(x) \ne c$, indicating the generator's ability to reveal model weaknesses~\cite{zhang2018lpips}.
\end{itemize}

\subsubsection{Procedure.}
For each dataset, we generate test inputs under the fixed and adaptive truncation schedules and apply the two probing methods (\emph{style-mixing} and \emph{truncation-only first-flip}). Each generated image undergoes automated screening and, if accepted, human verification.  
We then report (i) screening and validation rates per truncation level, (ii) diversity among validated inputs, (iii) fault detection ratios, and (iv) the relative efficiency of the two probing methods.  
Together, these measurements allow us to analyze how truncation affects the trade-off between \emph{validity}, \emph{diversity}, and \emph{fault revelation}, and to assess whether adaptive truncation or specific probing strategies offer systematic advantages.

\subsubsection{Human Validation.}
To obtain human validity, we conducted a small-scale annotation with two assessors. 
For each dataset and class, we prepared a folder containing generated flipped test cases and removed all metadata about their origin (e.g., truncation level $\psiT$, generation technique, and classifier outputs). 
Annotators therefore only saw images and their source classes. 
They were instructed to answer a binary question: \emph{``Is this image a valid example of this class?''} (yes/no). 
We marked an image as \emph{human-valid} if the annotator answered ''yes'' to the question.
Disagreements were resolved through discussion until consensus was reached.

\subsection{Effectiveness (RQ\textsubscript{1})} \label{sec:truncation-RQ1}

\autoref{tab:full-results} presents the results for all datasets and compared techniques. The results are averaged over the 25 runs.
Regarding \datasetA, without regularization ($\psiT{=}1.0$), the human-validated rate is \mbox{44\%} (11/25) with a cost of \mbox{42.00} seeds per validated case. As $\psiT$ decreases, validation improves and effort drops, peaking at \mbox{$\psiT{=}0.6$} with 80\% human-validated pairs. Pushing regularization further to \mbox{$\psiT{=}0.5$} slightly reduces validation (\mbox{68\%}). Similarly, in \datasetC results of human validation is highest at \mbox{$\psiT{=}0.6$} and slightly decreases at \mbox{$\psiT{=}0.5$}, while in \datasetB human validation peaks at \mbox{$\psiT{=}0.5$}.

\begin{figure}[t]
  \centering
  \includegraphics[width=0.9\columnwidth]{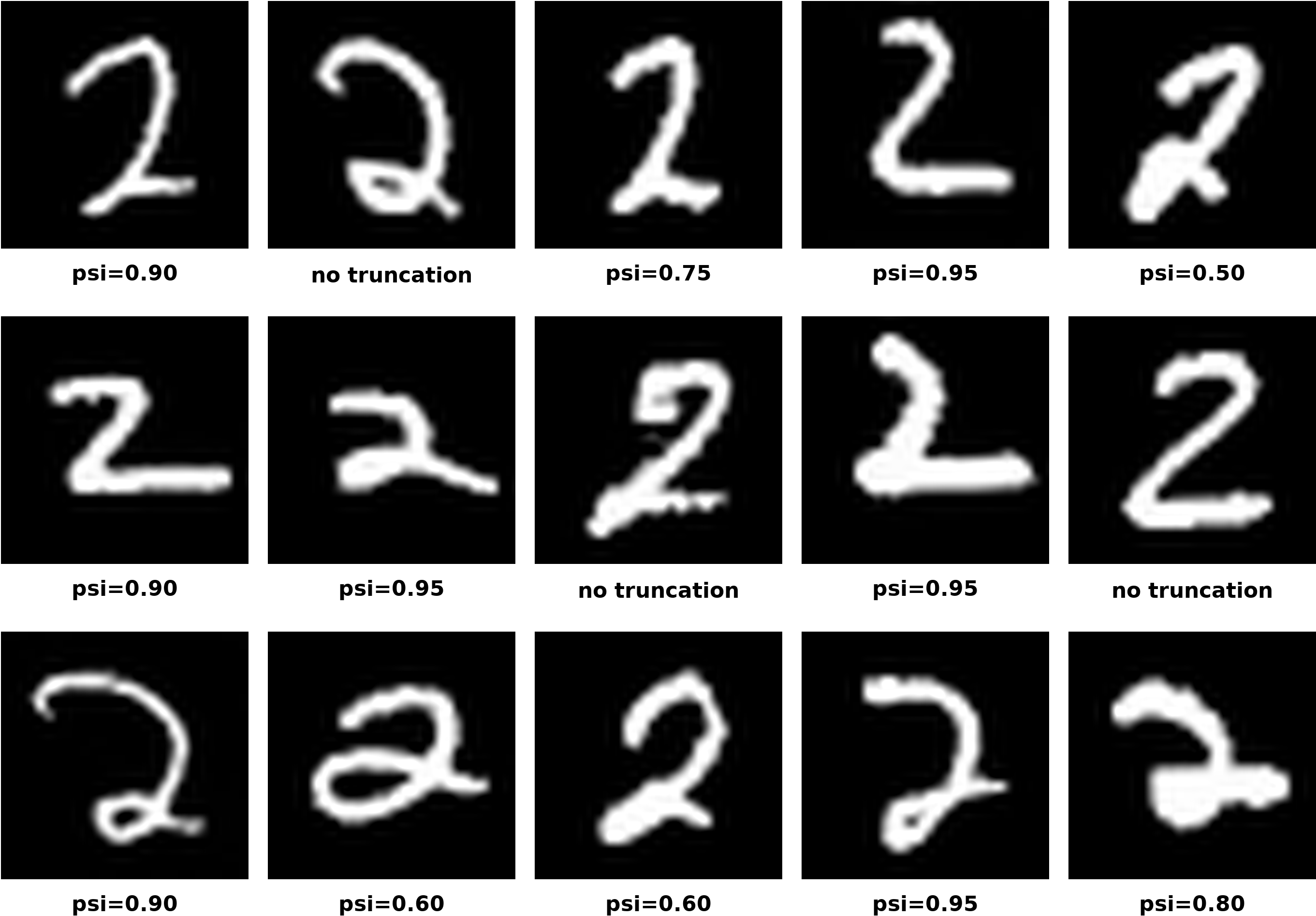}
  \caption{\datasetA: adaptive truncation, human-validated, fault-revealing test cases.}
  \label{fig:mnist_adaptive_grid}
\end{figure}

The \emph{adaptive} strategy achieves the best efficiency: it reaches 15 human-validated pairs using only 161 seeds in total, less than half of the effort required than at \mbox{$\psiT{=}0.6$}. This confirms the practical value of per-seed regularization: instead of discarding borderline seeds at $\psiT{=}1.0$, adaptively ``salvaging'' them near $\psistar$ yields more fault-prone tests at lower cost (see \autoref{fig:mnist_adaptive_grid}). We observed a similar result in \datasetB, where seed usage dropped from 526 (highest human validation) at \mbox{$\psiT{=}0.6$} to 161. While the adaptive strategy also decreased seed usage in \datasetC, the reduction was not as significant as in \datasetA and \datasetB.

Although \autoref{tab:full-results} primarily reports acceptance/efficiency, qualitative inspection in \autoref{fig:mnist_adaptive_grid} shows that \mbox{$\psiT{\approx}0.6$} maintains sufficient visual spread to reveal decision-boundary behavior. Extremely low $\psiT$ would risk collapsing diversity; our results suggest that mild to moderate regularization (and especially the adaptive strategy) strikes a better balance between human-validated realism and the variety needed to expose faults. Similar reasoning applies to the other datasets. \\

\rqbox{
\textbf{RQ\textsubscript{1} (effectiveness).} 
Mild–moderate latent regularization (\mbox{$\psiT{\approx}0.6$}) maximizes human-validated frontier yield on \datasetA. The adaptive policy minimizes human effort (seeds/valid) by salvaging hard seeds. Qualitative analysis shows that diversity remains adequate at these settings to expose faults, whereas overly strong regularization offers lower returns.
}

\subsection{Comparison (RQ\textsubscript{2})}\label{sec:rq5}

\begin{figure}[t]
  \centering
  \includegraphics[width=0.9\columnwidth]{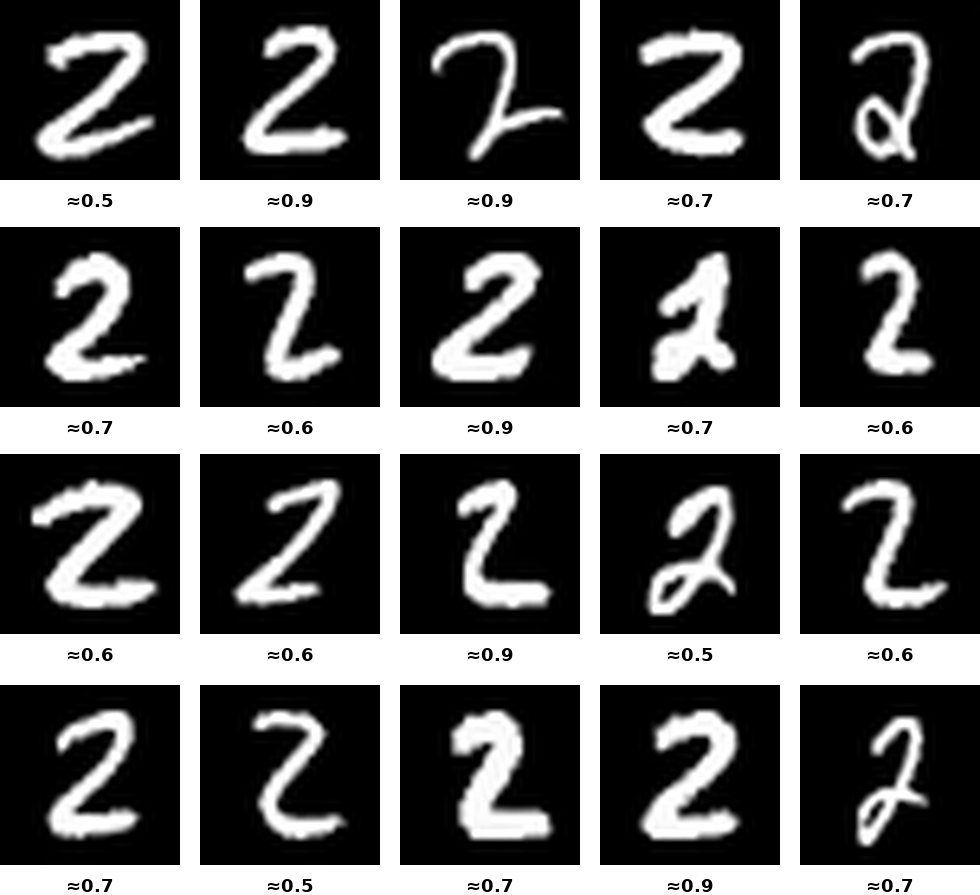}
  \caption{Cases from \datasetA where the first flip occurs at $\psistar$.}
  \label{fig:firstflip_grid}
\end{figure}

The two strategies appear to be complementary. \emph{First-flip} is lightweight and effective for quickly revealing faults under a truncation schedule (see \autoref{fig:firstflip_grid}); \emph{style-mixing} provides semantically guided boundary traversal with finer control, at the cost of extra search (see \autoref{fig:mnist_adaptive_grid}).

Across the same truncation settings used for RQ\textsubscript{1}, \emph{truncation-only first-flip} requires fewer design choices (no rival selection or layer sets) and rapidly identifies misbehaviours by descending $\psiT$. This makes it a strong configuration when computing time is limited. Style-mixing, in contrast, steers the latent representation toward specific rival classes via layer-wise interpolation and binary search; this extra complexity typically yields cleaner, semantically interpretable boundary crossings and can improve human acceptance for borderline cases, but incurs additional tuning (layers $L$, step budget $T$, rival selection).

On \datasetA, runs that reached the fixed quota of 25 frontiers showed that \emph{truncation-only first-flip} benefited markedly from adaptive seed salvage (lower seeds/valid) and produced a higher fraction of validated faults in our MNIST experiments, while \emph{style-mixing} produced boundary examples with more controllable semantics. 

On \datasetB, truncation-only first-flip achieves the lowest seed usage (93 seeds for 25 validated faults) and the highest fault detection rate, making it the most efficient option. In contrast, \emph{style-mixing} configurations require more seeds but allow finer control over semantic edits. 
 
On \datasetC, the distinction between the two techniques is less significant. Truncation-only first-flip remains competitive in fault detection, but its advantage over \emph{style-mixing} is smaller than on \datasetB, indicating that dataset characteristics influence how strongly each technique benefits from the first-flip technique. 
In short, if the goal is \emph{fast fault surfacing}, first-flip under adaptive $\psiT$ is preferable; if the goal includes \emph{interpretable semantics} for analysis and debugging, \emph{style-mixing} provides control at a moderate additional cost. \\

\rqbox{
\textbf{RQ\textsubscript{2} (comparison).} 
First-flip is the most efficient for fault discovery under latent regularization; style-mixing complements it by adding semantic control over boundary direction. 
}

\begin{figure*}[t]
  \centering
  \setlength{\tabcolsep}{1pt}
  \renewcommand{\arraystretch}{0}
  \scriptsize
  \resizebox{\textwidth}{!}{%
  \begin{tabular}{c c c c c c c c c c c}
    & $\psiT{=}1.0$ & $0.9$ & $0.8$ & $0.7$ & $0.6$ & $0.5$ & $0.4$ & $0.3$ & $0.2$ & $0.1$ \\[2pt]
    \rotatebox{90}{\ Up to 2nd layer} &
    \includegraphics[width=0.08\textwidth]{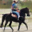} &
    \includegraphics[width=0.08\textwidth]{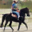} &
    \includegraphics[width=0.08\textwidth]{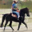} &
    \includegraphics[width=0.08\textwidth]{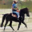} &
    \includegraphics[width=0.08\textwidth]{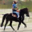} &
    \includegraphics[width=0.08\textwidth]{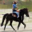} &
    \includegraphics[width=0.08\textwidth]{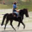} &
    \includegraphics[width=0.08\textwidth]{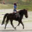} &
    \includegraphics[width=0.08\textwidth]{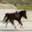} &
    \includegraphics[width=0.08\textwidth]{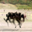} \\[4pt]

    \rotatebox{90}{\ Up to 5th layer} &
    \includegraphics[width=0.08\textwidth]{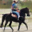} &
    \includegraphics[width=0.08\textwidth]{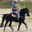} &
    \includegraphics[width=0.08\textwidth]{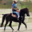} &
    \includegraphics[width=0.08\textwidth]{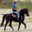} &
    \includegraphics[width=0.08\textwidth]{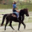} &
    \includegraphics[width=0.08\textwidth]{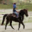} &
    \includegraphics[width=0.08\textwidth]{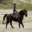} &
    \includegraphics[width=0.08\textwidth]{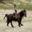} &
    \includegraphics[width=0.08\textwidth]{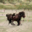} &
    \includegraphics[width=0.08\textwidth]{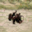} \\[4pt]

    \rotatebox{90}{\ Up to 8th layer} &
    \includegraphics[width=0.08\textwidth]{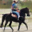} &
    \includegraphics[width=0.08\textwidth]{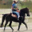} &
    \includegraphics[width=0.08\textwidth]{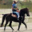} &
    \includegraphics[width=0.08\textwidth]{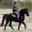} &
    \includegraphics[width=0.08\textwidth]{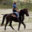} &
    \includegraphics[width=0.08\textwidth]{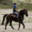} &
    \includegraphics[width=0.08\textwidth]{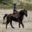} &
    \includegraphics[width=0.08\textwidth]{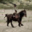} &
    \includegraphics[width=0.08\textwidth]{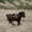} &
    \includegraphics[width=0.08\textwidth]{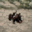} \\
  \end{tabular}}

  \caption{\textbf{Layer-specific truncation sweeps (same seed).} Columns show decreasing $\psi$ from 1.0 to 0.1; rows indicate which layers are truncated: up to layer 2 (coarse), up to layer 5 (coarse+middle), and up to layer 8 (all layers).
}
  \label{fig:trunc_cutoff}
\end{figure*}

\subsection{Threats to Validity}
\subsubsection{Internal validity}
A potential threat arises from the choice of latent seeds, as some seeds may inherently yield more realistic or diverse samples. We mitigate this by reusing identical seed lists across all truncation budgets and probing methods. Strong truncation may also induce mode collapse, reducing diversity and artificially inflating validity; to limit this effect, we restrict truncation sweeps to $\psiT \ge 0.5$. In addition, the degree of latent disentanglement varies across datasets and classes, which can make the effects of truncation or layer-wise edits less consistent. This reflects an intrinsic limitation of current generative models rather than an artifact of our setup.

Our human-validity labels are based on a small-scale annotation with two assessors and therefore introduce subjectivity. We treat this step as a lightweight plausibility filter and interpret it jointly with automated screening, diversity, and fault-detection metrics. Similarly, the \lpips metric used for filtering is only an approximation of semantic similarity, but it provides a consistent and reproducible basis for comparing truncation levels and probing strategies.

\subsubsection{External validity}

Experiments are conducted on small-scale image datasets with relatively simple classifiers, and the observed relationships between truncation, validity, and diversity may not directly transfer to larger or more complex domains. Moreover, our conclusions apply specifically to StyleGAN-based generators; truncation may behave differently in other architectures, such as diffusion or transformer-based models. Extending this analysis to additional modalities and generative paradigms is an important direction for future work.

To ensure transparency and replicability, we release all experimental scripts (including seed lists and fixed-noise configurations), JSON logs, and plotting templates required to reproduce every figure and table in our replication package~\cite{replication-package}.

\section{Discussion}

Our results show that truncation can serve as a simple yet effective control for enhancing the quality of generated test inputs in latent-space testing frameworks such as \textsc{Mimicry}. Moderate truncation levels (\mbox{$\psiT \approx 0.6$}) tend to maximize the fraction of human-validated valid and fault-revealing test cases, striking a balance between semantic realism and diversity.  

The adaptive truncation mechanism further improves efficiency by salvaging seeds that would otherwise be discarded, leading to a richer and more fault-prone test pool. This adaptivity makes truncation-based refinement particularly suitable when annotation or computational budgets are limited.  

Between the two probing methods, \emph{style-mixing} provides structured, semantically meaningful perturbations but requires additional computation and hyperparameter tuning. In contrast, the \emph{truncation-only first-flip} approach offers a lightweight, diagnostic alternative that can reveal classifier weaknesses with minimal setup. Combining both techniques allows a comprehensive exploration of decision boundaries, from coarse latent refinements to fine-grained semantic traversals.

In addition to $\psiT$, StyleGAN exposes adjacent controls that plausibly affect test case validity and diversity: (i) per-layer truncation cutoffs that restrict where truncation is applied in $\Wplus$, (ii) the magnitude/schedule of stochastic noise injection in the synthesis network, and (iii) the choice of layers used for style mixing~\cite{karras2019stylegan}. 

\section{Conclusion and Future Work}\label{sec:conclusion}

This study investigated how truncation—a common latent-space regularization technique—affects the generation of test inputs for deep learning classifiers. We quantified its impact on human-verified validity, diversity, and fault detection, and introduced a simple yet effective \emph{first-flip} probe based solely on truncation. We further compared this approach with a truncation-assisted \emph{style-mixing} probe that explores the latent space through semantic interpolation.

Our findings show that moderate truncation levels and adaptive seed refinement yield the best balance between validity, diversity, and fault revelation. These results provide practical guidance and actionable defaults for configuring generative test input generators, as well as a reproducible protocol to support future research in deep learning testing.

In our experiments, we focus on StyleGAN, but similar fidelity controls appear in other generative architectures, such as guidance in diffusion models and temperature, top-$k$, and top-$p$ in autoregressive decoders. The same sample-and-screen design can be reused to calibrate these knobs in diffusion or transformer-based generators to balance the fidelity–diversity trade-off. We leave validation of this generalization to future work.

Future extensions of this study could explore truncation effects across StyleGAN's layers as shown in \autoref{fig:trunc_cutoff}, as well as related generator-level controls such as synthesis noise or layer-selection choices, and in more complex generative architectures, such as diffusion or transformer-based models, where latent representations and fidelity controls differ fundamentally from GANs. Finally, integrating automated validity estimation and adaptive truncation scheduling could reduce reliance on human annotation, advancing scalable and self-adaptive test generation for AI-based systems.

\balance
\bibliographystyle{ACM-Reference-Format}
\bibliography{references.bib}
\end{document}